# MAGNETIZATION AND MAGNETOTRANSPORT OF $LnBaCo_2O_{5.5}$ (Ln=Gd, Eu) SINGLE CRYSTALS

Z.X. Zhou, S. McCall*, C.S. Alexander, J.E. Crow, and P. Schlottmann**
National High Magnetic Field Laboratory, Tallahassee, FL 32310

S.N. Barilo, S.V. Shiryaev and G.L Bychkov
Institute of Solid State and Semiconductor Physics
Academy of Science, Minsk, Belarus

and

R.P. Guertin
Department of Physics, Tufts University, Boston, MA 02155

**Abstract**

The magnetization, resistivity and magnetoresistance (MR) of single crystals of $GdBaCo_2O_{5.5}$ and $EuBaCo_2O_{5.5}$ are measured over a wide range of dc magnetic fields (up to 30 T) and temperature. In $LnBaCo_2O_{5.5}$ (Ln=Gd, Eu), the Co-ions are trivalent and can exist in three spin states, namely, the S=0 low spin state (LS), the S=1 intermediate spin state (IS) and the S=2 high spin state (HS). We confirm that $GdBaCo_2O_{5.5}$ and $EuBaCo_2O_{5.5}$ have a metal-insulator transition accompanied by a spin-state transition at $T_{MI} \approx 365$ and 335 K, respectively. The data suggest an equal ratio of LS (S=0) and IS (S=1) $Co^{3+}$ ions below $T_{MI}$, with no indication of additional spin state transitions. The low field magnetization shows a transition to a highly anisotropic ferromagnetic phase at 270 K, followed by another magnetic transition to an antiferromagnetic phase at a slightly lower temperature. The magnetization data are suggestive of weak correlations between the Gd-spins but no clear signature of ordering is seen for T > 2 K. Significant anisotropy between the a-b plane and c axis was observed in magnetic and magnetotransport properties for both compounds. For $GdBaCo_2O_{5.5}$, the resistivity and MR data imply a strong correlation between the spin-order and charge carriers. For $EuBaCo_2O_{5.5}$, the magnetic phase diagram is very similar to its Gd counterpart, but the low-T MR with current flow in the ab plane is positive rather than negative as for Gd. The magnitude and the hysteresis of the MR for $EuBaCo_2O_{5.5}$ decrease with increasing temperature, and at higher T the MR changes sign and becomes negative. The difference in the behavior of both compounds may arise from a small valence admixture in the nonmagnetic Eu ions, i.e. a valence slightly less than 3+.

**Introduction:**

The existence of several possible spin states in a given oxidation state makes the cobaltites a rich but also challenging system to study. $Co^{3+}$ has three possible spin states: the low spin state (LS, $t_{2g}^6 e_g^0$, S=0), the intermediate spin state (IS, $t_{2g}^5 e_g^1$, S=1), and the high spin state (HS, $t_{2g}^4 e_g^2$, S=2) which arise from the competition between crystal field (CF), on-site Coulomb correlations, and the intra-atomic exchange energy.[1] As the simplest representative of trivalent perovskite cobalt oxides, $LaCoO_3$ has been investigated extensively. These studies demonstrated the complex magnetism among the Co ions and the difficulty in determining the spin state of $Co^{3+}$, stimulating interest in searching for other cobaltites.[2-4] The $LnBaCo_2O_{5+\delta}$ (Ln = rare earth) series, in particular, displays a variety of interesting phenomena, including metal-insulator transitions, spin state ordering and a giant magnetoresistance, which depend on the valence and spin states of the Co ions.[5-14] The valence state of the Co ions is determined by the oxygen content $\delta$ ($0 \leq \delta \leq 1$), where all the Co ions are expected to be trivalent when $\delta=0.5$. For the case $\delta=0.5$, one half of the cobalt occupy octahedral sites ($CoO_6$) while the other half possess square pyramidal symmetry ($CoO_5$), forming alternating oxygen rich and oxygen deficient a-c layers along the b axis (e.g. see the inset of Fig. 1 in Ref. 1). For $0.0<\delta<0.5$ and $0.5<\delta<1.0$, divalent and tetravalent Co ions coexist with trivalent Co ions, respectively. In the special cases of $\delta=0$ and $\delta=1$, the cobalt environment appears to be exclusively pyramidal ($\delta=0$) or octahedral ($\delta=1$).

A metal-insulator transition was observed in $GdBaCo_2O_{5.5}$ and $TbBaCo_2O_{5.5}$ and an anomalous change of lattice constants has been reported at the metal-insulator transition temperature indicating that the transition is structure related.[7, 15] Curie Weiss fits of the magnetic susceptibility for $GdBaCo_2O_{5.5}$ and $TbBaCo_2O_{5.5}$ show a drastic change in the effective moment ($\mu_{eff}$) across $T_{MI}$, suggesting that the metal-insulator transition is also associated with a spin-state transition.[1, 7, 10] Similar results for $NdBaCo_2O_{5.5}$,[12] suggest that this is a general trend in this cobaltite series. The ultrahigh resolution synchrotron diffraction data reported by C. Frontera *et al*[1] show that the average Co-O bond distance within the octahedra increases substantially at $T_{MI}$ upon heating, while it decreases in the pyramids, implying that the spin-state transition in $LnBaCo_2O_{5.5}$ occurs in the octahedra. This suggests that the Co ions in pyramids have IS (S=1) and those in the octahedra have LS (S=0) below $T_{MI}$. It is commonly observed in these systems that there is ferromagnetism (FM) in a narrow temperature region sandwiched between a high temperature paramagnetic state and a low temperature antiferromagnetic (AFM) state.

At least two different models have been proposed to explain the narrow temperature region of FM order and the subsequent FM→AFM transition upon cooling. (1) Moritomo *et al*[7] suggest that the ferromagnetism and FM→AFM transition could arise from the competition between an inherent AFM superexchange interaction and a FM double-exchange interaction due to carrier localization. (2) Roy et al proposed two sublattices facilitating an in-plane FM interaction and an inter-plane AFM interaction.[10] This second model was further elaborated independently by Taskin *et al*[13] and Khalyvin *et al*[14] based on their magnetization data measured on $GdBaCo_2O_{5.5}$ single crystals. Our magnetization data suggest a two-sublattice model similar to the latter with temperature dependent inter-plane coupling, which might be caused by a competition between an AFM superexchange interaction and a FM double-exchange interaction. In order to further investigate the nature of the magnetism in this system, a detailed study of the magnetic anisotropy is required. When this study began, all the published studies used polycrystalline samples and the lack of single crystal data made it impossible to probe the anisotropy via magnetic and transport measurements for any $LnBaCo_2O_{5.5}$ compound. We have produced single crystal samples of $GdBaCo_2O_{5+\delta}$ (in parallel with two other groups[13, 14]) and $EuBaCo_2O_{5+\delta}$ allowing a study of the anisotropy. Our results indicate very large anisotropy in the magnetization, M(H), between the a-b plane and c axis. High field isothermal magnetization measurements performed on single crystal $GdBaCo_2O_{5.5}$ enable accurate determination of the saturation moment and lead to a spin model for the magnetically ordered $Co^{3+}$ in the pyramidal sites.

**Experiment:**

Crystals of $GdBaCo_2O_{5+\delta}$ and $EuBaCo_2O_{5+\delta}$ were grown from a high temperature flux melt using an off-stoichiometric initial mixture of the corresponding oxides. The mixtures were placed into $Al_2O_3$ crucibles and heated to 1350 °C. After homogenization for 2 hours, the flux melt was slowly cooled to 1000 °C at the rate of 1 °C/hr and then cooled quickly to room temperature. The as-grown crystals have tetragonal crystal symmetry (space group P4/mmm) as determined by X-ray diffraction. The as-grown crystals were then annealed for several days at 600 °C in 3 bar extra oxygen followed by cooling to room temperature at a rate of 10 °C/hr. During the cooling, the crystal symmetry changes from tetragonal to orthorhombic (space group Pmmm). Due to this transition, the resulting crystals display a twinned structure in which the a and b axes are intermixed as determined by Laue diffraction. Thermogravimetric analysis has indicated that the mobile oxygen atoms located within the Gd-O plane, can be reversibly added and removed from the system in the temperature range 400-600 K. Crystal phase purity and cation composition of the crystals were checked by x-ray diffraction and x-ray fluorescent analysis. The oxygen content in the crystals was determined by iodometric titration to be 5.47 ± 0.02 for $GdBaCo_2O_{5+\delta}$. Using iodometric titration, we also determined δ = 0.47 ± 0.01 for the $EuBaCo_2O_{5+\delta}$ crystal after the same oxygenation procedure. Magnetization data below 7 T were taken in a Quantum Design SQUID magnetometer, while a vibrating sample magnetometer (VSM) was used to measure isothermal magnetization up to 30 T in continuous magnetic fields at the National High Magnetic Field Laboratory in Tallahassee, Florida. Resistivity and magnetoresistance were measured using a standard four-probe technique. Gold contacts were sputtered onto the samples to minimize contact resistance, which was prohibitively high without the gold, especially at low temperatures.

**Results:**

The magnetization, M(T) for $GdBaCo_2O_{5.5}$ upon cooling with H ⊥ c increases rapidly at 270 K reaching a maximum around 250 K, as shown in Fig. 1(a). Subsequently, M(T) drops abruptly, leveling off at 210 K. The sharp increase of M(T) at 270 K has been interpreted as the onset of ferromagnetic ordering with the subsequent sudden drop indicating a FM→AFM transition. A much smaller and broader peak appears near 150 K. That feature was also observed as a change of slope in the thermal expansion measurement reported by Roy *et al* who interpreted it as a second order phase transition to a second magnetic state (within the IS state).[10] The Curie-like paramagnetic M(T) for T<100 K (low temperature up-turn) is attributed to the Gd spins. Our attempts to interpret this component of M(T) as paramagnetic Gd spins were only partially successful, even when a simple antiferromagnetic correlation between Gd-spins (negative Weiss-temperature) was included. Therefore, there must be a weak interaction between the Gd and Co spins.

In contrast, M(T) for H || c is predominantly paramagnetic except for a small feature near 250 K. The M(T) curves for H ⊥ c and H || c demonstrate the strong anisotropic magnetic characteristics with the easy axis in the a-b plane. The small increase near 250 K could arise from a small component of the magnetization in the a-b plane due to a slight misalignment of the sample. The height of this peak near 250 K is about 50 times larger for H ⊥ c than for H || c, which could be explained by an ~2° misalignment between the c-axis and the applied field. The Curie-Weiss tail due to the Gd ions for T<100 K is somewhat weaker for H || c than for H ⊥ c. Perhaps a weaker Co-generated molecular field exists at the Gd sites when the external field is applied parallel to the c-axis (hard Co magnetization axis).

For comparison, we show in Fig. 1(b) the zero field cooled (ZFC) and field cooled (FC) M(T) for $EuBaCo_2O_{5.5}$. Unlike for $GdBaCo_2O_{5.5}$, a significant discrepancy exists between the ZFC and FC M(T) curves for $EuBaCo_2O_{5.5}$. As for $GdBaCo_2O_{5.5}$, a sharp peak appears in M(T) at 210 K for $EuBaCo_2O_{5.5}$, but its magnitude is an order of magnitude smaller than for the Gd counterpart for H ⊥ c. The ratio of the peak height with H ⊥ c to that with H || c is about 10, significantly smaller than the same ratio in $GdBaCo_2O_{5.5}$. Therefore, it is difficult in this case to attribute the peak along c-axis to a misalignment of the sample. In contrast to $Gd^{3+}$, the $Eu^{3+}$-ions do not have a magnetic moment, so the data lack an upturn in M(T) at low temperature. However, a small but significant feature can be observed below 45 K in the Eu sample, which is barely discernable in the Gd sample, even if the Gd contribution is subtracted using a Curie law for free $Gd^{3+}$ ions. The feature consists of an increase in the FC magnetization and a peak in the ZFC magnetization, observed for both H ⊥ c and for H ∥ c. The difference between $M_{ZFC}$ and $M_{FC}$ could originate from a Eu-Co interaction arising from a probable valence admixture in the Eu-ions (valence less than but close to 3+). This interaction could result in a frustration of magnetic order. Our data suggest that AFM long-range order coexist with spin-glass-type features in $EuBaCo_2O_{5.5}$ below 45 K. Spin-glass-type features could, however, also arise due to inhomogeneous oxygenation.

There are several Eu-based intermetallic compounds exhibiting intermediate valence. A small valence admixture of $Eu^{2+}$ is also expected from the general trend of the lattice parameters and the unit cell volume for the rare earth (Ln) series $LnBaCo_2O_{5+\delta}$ (see table I and Fig. 6a of Ref. 6). As compared to the interpolation between the trivalent Sm and Gd compounds, the a- and b-lattice parameters for Eu are slightly larger, while the c-lattice parameter is somewhat smaller. The volume of the unit cell of the Eu compound is slightly larger than the interpolation. A divalent admixture in the Eu-compound is then plausible, since $Eu^{2+}$ has a larger ionic radius. The f-shell of $Eu^{2+}$ is isoelectronic to $Gd^{3+}$ and is magnetic. Hence, a small valence admixture would give rise to weak magnetic contributions arising from the Eu ions.

The isothermal magnetization of $GdBaCo_2O_{5.5}$ was measured at several temperatures in dc fields up to 30T applied both parallel and perpendicular to the c-axis. Three features of the T=5.0 K magnetization data in Fig. 2 should be noted: (1) The approach to magnetic saturation, which is dominated by the $Gd^{3+}$ ions, is more rapid for H ⊥ c than for H ∥ c, again demonstrating the magnetic anisotropy in these single crystals. (2) A field-induced transition arises in the magnetization for H ⊥ c at 5 K and 6-10 T, while no similar transition appears for H ∥ c up to 30 T. (3) Hysteresis, evident at low fields, is significant for H ⊥ c and barely discernable for H ∥ c. The hysteresis decreases with temperature, but persists up to about 260 K.

Assuming a Brillouin-like magnetization for the Gd ions at 5 K, the Gd moments should be fully polarized by an external field of 30 T. The relative flatness of M(H) at high fields demonstrates that the moments of the Co ions are also saturated. Within the experimental error of the VSM measurement, the saturation moment obtained from the M(H) curve for H ⊥ c at 1.8 K (data not shown) is $0.88\mu_B$/Co. This is consistent with a 50%-50% mixture of LS (S=0) and IS (S=1) states ($1\mu_B$/Co, assuming complete quenching of the orbital angular momentum with the unpaired spins all aligned).

The hysteretic transition discussed above is attributed to the $Co^{3+}$-spins and shows the strong anisotropy of the magnetization between a field applied in the a-b plane and along the c-axis. The broad transition in the hysteresis loop observed for H $\perp$ c is probably the consequence of a spin flip (AFM $\rightarrow$ FM) and detwinning of magnetic domains in the a-b plane due to the large magnetic field. In zero field, the crystal is twinned in the a-b plane, and with increasing magnetic field in the a-b plane the anisotropy energy and the magnetic energy compete, rotating the magnetic domains with spins initially perpendicular to the field towards the field direction. Eventually, the magnetic energy outweighs the anisotropy energy and all the magnetic domains become aligned along the magnetic field.

The field at which the spin flip transition occurs decreases with increasing temperature and approaches zero at 260 K. Above 78 K, there is no sign of detwinning the magnetization of the $GdBaCo_2O_{5.5}$ crystal up to fields of 7 T. Some representative M(H) curves for H $\perp$ and $\parallel$ c are displayed in Fig. 3(a) for T=170 K, which also show how the Gd background can be subtracted out. After subtracting the linear contribution (note that the Brillouin function is essentially linear for high temperatures) to the magnetization due to paramagnetic Gd, M(H$\perp$c), appears to saturate at a relatively low field (H < 3T) for 150 < T < 260 K (inset of Fig. 3(a)), while M (H $\parallel$ c) exhibits linear behavior to 7 T (Fig. 3(a)), indicating that the field slightly tilts the moment away from the a-b plane. In Fig. 3(b), we show a quite similar behavior at T = 78 K.

As shown in the inset of Fig. 3(b), the saturation moment extracted from the M(H$\perp$c) curves between 78 < T < 260 K fits well to a $T^3$-dependence returning a moment of $0.49\mu_B$/Co when extrapolated to T=0 K. This corresponds to only half of the value expected assuming half of the $Co^{3+}$-ions are in the IS state (S=1) and the other half are in the LS state (S=0). Due to the extensive twinning in the a-b plane, for H $\parallel$ a or H $\parallel$ b, half of the intermediate spins are parallel to the applied field while the other half are perpendicular to it, which then leads to an expected saturation moment of $0.5\mu_B$/Co as observed. If the field is applied along an arbitrary direction within the a-b plane, it may saturate the spins along both the a-axis and b-axis leading to a resultant moment of 0.5-0.71$\mu_B$/Co with the maximum corresponding to a 45 degree angle between the field and the a- or b-axis. The extracted saturation moment may not necessarily follow the $T^3$-dependence below 78K. However, as a rough estimate it agrees quite well with an equal ratio of LS and IS states and twinning in the a-b plane. The $T^3$-dependence could arise from spin-waves with linear long wavelength dispersion due to the coupling between the planes.

For comparison we present in Fig. 3(c) the M(H) for $EuBaCo_2O_{5.5}$ at 5 K in dc fields up to 30 T applied both parallel and perpendicular to the c axis. At low fields significant hysteresis is observed for H $\perp$ c, while the magnetization for H $\parallel$ c exhibits a linear field dependence up to 30 T with no sign of hysteresis, again demonstrating the significant magnetic anisotropy in $EuBaCo_2O_{5.5}$ with the c-axis being the hard axis. Unlike $GdBaCo_2O_{5.5}$, the M(H) of $EuBaCo_2O_{5.5}$ has no indication of saturation up to 30 T. The underlying linear magnetization is likely to arise from $Co^{3+}$ ions and from the Eu ions, which have a slight valence admixture (thus also contribute to the magnetization). The smaller moment of the $Co^{3+}$ ions in $EuBaCo_2O_{5.5}$ as compared to $GdBaCo_2O_{5.5}$ is also consistent with the magnetization shown in Figs. 1(a) and (b).

The resistivity data for $GdBaCo_2O_{5.5}$ also show an important anisotropy between the a-b plane and c-axis as shown in Fig. 4(a). In zero-magnetic field, the $\rho(T)$ in the a-b plane breaks up into regions consistent with the magnetization data for 110 < T < 294 K, implying strong coupling between the spin-order and charge carriers (see Fig. 4(b)). On the other hand, the resistivity along the c-axis stays virtually flat above 165 K and then rises abruptly when the temperature is further decreased (see Fig. 4(a)). The low temperature resistivity data (T < 110 K) measured both in the a-b plane and along the c axis fit well to the expression for the 3D variable-range hopping (VRH) model, $\rho=\rho_0 \exp(T_0/T)^{1/4}$, as shown in Fig. 4(c). At low temperatures (T < 110K), the promotion of d-electrons into the conduction band is frozen out and VRH of localized electrons is the only mechanism of conduction. The disorder or randomness of the potential that is necessary for the VRH to occur is likely due to the defects created by slight oxygen deficiency. When the temperature is increased, the d-electrons are thermally excited to the conduction band and eventually an exponential activation law takes over. Since the magnetic ordering of the spins affects the gap between the conduction band and valence band, it is not surprising that the a-b plane resistivity breaks up into regions consistent with the magnetization data for 110 < T < 294 K, where a simple activation is the dominant mechanism of conduction (see Fig 4(b)).

The resistivity of $EuBaCo_2O_{5.5}$ measured with current flow in the ab plane and along the c-axis is shown in Fig. 5. A substantial anisotropy is again present, but the behavior is qualitatively different from the Gd counterpart. The $\rho(T)$ for I $\perp$ c increases slowly with decreasing temperature until T = 50 K and below this T it begins to rise rapidly. Note that from 300 K to 50 K, $\rho(T)$ only increases by a factor of six, while it increases by about seven orders of magnitude from 50 K down to 4 K. This demonstrates that different conduction mechanisms are relevant for the temperature regions above and below 50 K. The resistivity data (I $\perp$ c) for $EuBaCo_2O_{5.5}$ below 50 K can be fit with the 3D VRH model mentioned above. The resistivity along the c-axis smoothly increases with decreasing temperature down to the lowest T measured. The data for T < 120 K can again be fit to the expression of the VRH model as shown in the inset of Fig. 5.

The transverse magnetoresistivity (I in the a-b plane and $\perp$ H), $\rho(H)$, for $GdBaCo_2O_{5.5}$ and $EuBaCo_2O_{5.5}$ was measured for H $\parallel$ and $\perp$ c. For $GdBaCo_2O_{5.5}$ there is considerable anisotropy in the isothermal magnetoresistance (MR) with regard to the field direction. A hysteretic giant negative MR takes place for H $\perp$ c. The large decrease of the MR is correlated with the field induced transition in the magnetization. The relative MR defined as MR (%)=100×($\rho(H)$- $\rho(0)$)/ $\rho(0)$ is shown in Fig. 6(a), and again demonstrates a strong anisotropy of -93% for H $\perp$ c and -22% for H $\parallel$ c at 40 K and 17 T.[11] The giant negative MR increases monotonically with decreasing temperature and is most pronounced in the VRH temperature range (T <110 K). This can be understood in terms of a shift in the mobility edge. In the VRH temperature regime, all states with energy less than the mobility edge are localized while states with higher energy are non-localized or extended. If the Fermi level is moved across the mobility edge, a dramatic change in resistivity is achieved.[17] In zero field the Fermi energy $E_F$ lies below the mobility edge leading to the observed VRH conduction. When a sufficiently large field is applied in the ab plane, the conduction band splits into two spin subbands causing the mobility edge of each subband to shift up and down, respectively. As a result, the mobility edge of one spin subband may drop below the Fermi level leading to a drastic increase in the conductivity.

As shown in Fig. 6(b), EuBaCo$_2$O$_{5.5}$ has positive MR, in contrast to its Gd counterpart. At low T, the MR is large and hysteretic, which is consistent with the magnetization data. With increasing temperature the MR is gradually reduced and at 160 K the MR is already negative (see Fig. 6(c)). Hence, the MR for I $\perp$ c, H $\perp$ c and H $\perp$ I changes sign as a function of temperature. The MR of EuBaCo$_2$O$_{5.5}$ for all T strongly depends on the relative directions of I, H and the crystal c-axis, as shown in Fig. 6(c) for T=160 K. The differences in the MR of GdBaCo$_2$O$_{5.5}$ and EuBaCo$_2$O$_{5.5}$ are hard to understand in view of their similar crystal and magnetic structures. However, the key difference between the two systems is that while the Gd-ions are trivalent and have a large magnetic moment, the nonmagnetic Eu-ions probably have a small valence admixture from the energetically nearby Eu$^{2+}$ configuration yielding an intermediate valence slightly less than trivalent. Even a small admixture (e.g., 5 to 10%) would change the charge distribution as well as induce a weak magnetic polarizability at the Eu sites.

**Discussion:**

Below 100 K, the paramagnetic contribution from Gd dominates M(T) and our analysis suggests weak antiferromagnetic correlations between Gd sites corresponding to a paramagnetic Weiss temperature of a few Kelvin. In addition, there is a weak interaction between Gd and Co ions. However, the Curie-Weiss temperature, $\Theta$, associated with the ferromagnetic correlations between Co sites in the paramagnetic region (280 < T < 360 K) between the metal-insulator transition and the FM ordered state is above 200 K. To determine the effective moment of the Co, $\mu^{Co}_{eff}$, and $\Theta$, it is necessary to subtract the Gd contribution from the magnetic susceptibility. After subtracting the Gd contributions, a Curie-Weiss fit of $\chi$(T) for 280 < T < 350 K yields $\Theta$ = 265 K and $\mu^{Co}_{eff}$ = 2.1$\mu_B$/Co, which is close to the expected value of 2$\mu_B$/Co for an equal ratio of LS (S=0) and IS (S=1) Co$^{3+}$ ions based on the formula: $(\mu^{Co}_{eff})^2$ = 0.5×$(\mu^{Co}_{eff}$ (LS)$)^2$ +0.5×$(\mu^{Co}_{eff}$ (IS)$)^2$.

The extrapolated saturation moment of 0.49$\mu_B$/Co shown in the inset of Fig. 3(b) also supports a mixture of an equal number of IS and LS sites, assuming twinning within the a-b plane. The isothermal magnetization at T=1.8 K up to 30 T yields a saturation moment of 0.88 $\mu_B$/Co, also suggesting a 1:1 ratio of LS:IS in a sample with detwinned magnetic domains due to the high magnetic field. Therefore, our magnetization data are compatible with an equal ratio of LS and IS below $T_{MI}$ ~ 365 K, which remains unchanged at lower temperatures. There is no indication of a spin state transition below $T_{MI}$, in agreement with the ultrahigh resolution synchrotron x-ray powder diffraction (SXRPD) data for 10 K ≤ T ≤ 400 K, showing no structural change below $T_{MI}$.[15]

Magnetization, as a bulk probe, does not provide a microscopic picture that enables determination of the spin states for the pyramidal and octahedral sites. Combining our results with the SXRPD data reported by C. Frontera *et al*[1], we propose a two sublattice model, where the $Co^{3+}$-ions at the octahedral sites ($CoO_6$) are in a LS (S = 0) state and the $Co^{3+}$-ions at the pyramidal sites ($CoO_5$) are in the IS (S = 1) state, thus leading to alternating planes of IS $Co^{3+}$-ions separated by non-magnetic planes of LS $Co^{3+}$-ions. The IS sites form two-leg ladders with rungs made of two $CoO_5$ square pyramids bridged by an apical oxygen. These ladders order ferromagnetically within the planes while adjacent sets of $CoO_5$ planes weakly couple to each other with the sign of coupling depending on the temperature. This model is consistent with the magnetization study on detwinned $GdBaCo_2O_{5.5}$ crystals by Taskin *et al*[13], although it does not require an Ising-like spin anisotropy in the interactions. Below $T_N$ (~260 K),[10, 13, 14] the inter-plane coupling is antiferromagnetic leading to an overall AFM, which could be attributed to the superexchange between the $Co^{3+}$ moments mediated by two oxygen sites and one LS $Co^{3+}$-ion for each such pair of IS $Co^{3+}$-ions.[13, 16] A second interaction competing with the superexchange for dominance is needed to explain the temperature or field induced AFM→FM transition. We believe this competing interaction is a thermally activated double-exchange. With increasing temperature, the number of d-electrons excited to the conduction band increases, so that more electrons participate in the double-exchange via hopping in the conduction band. Thus with increasing T the double-exchange interaction is enhanced and can lead to the inter-plane AFM→FM transition. A magnetic field can also boost the double-exchange mediated by the conduction band by reducing the gap between the conduction and valence bands and aligning the spins. Indeed, a moderate magnetic field (< 3 T) is able to induce an AFM to FM transition, indicating that the AFM inter-plane ordering below $T_N$ is relatively weak and sensitive to the magnetic field. Simple thermal activation dependences were also found in the resistivity above 110 K.

In a crystal with twinning in the a-b plane, this model predicts a saturation moment of $0.5\mu_B$/Co when extrapolated to T=0 K. We assume that due to twinning 50% of Co spins in the IS state are ordered along the field direction and 50% are ordered perpendicular to this direction. This requires a field induced AFM→FM (spin-flip) transition below $T_N$, which aligns spins in the domains with order along the field direction, in good agreement with our experimental result of $0.49\mu_B$/Co. The same model also predicts a saturation moment of $1\mu_B$/Co in a field high enough to induce an AFM→FM transition and overcome the anisotropy energy between the a and b axes so that all the Co moments in the IS state (50% of the total Co ions) align along the field direction regardless of their initial orientations. This is consistent with the saturation moment obtained from our isothermal magnetization measurement at T = 1.8 K to 30 T. The above arguments are independent of the assignment of IS and LS to specific $Co^{3+}$ environments, and are still valid if we assign IS to the $Co^{3+}$ ions in the octahedral environment and LS to the pyramidal sites.

The inverse susceptibility of $EuBaCo_2O_{5.5}$ above $T_c$ breaks into two distinct paramagnetic regions at $T_{MI} \approx 335$ K (data not shown), suggesting a possible spin state transition associated with a metal insulator transition. A similar Curie-Weiss fit to the $\chi(T)$ data for 245 < T < 325 K gives a ferromagnetic Curie-Weiss temperature, $\Theta$ = 222 K, and $\mu_{eff}$ =2.42$\mu_B$/Co, which is significantly larger than the effective moment expected for $EuBaCo_2O_{5.5}$ ($2\mu_B$/Co) if all its Eu-ions were trivalent. This difference between the experimental and calculated values suggests the existence of some valence admixture in the Eu ions (effective valence less than 3+), which we believe is also the cause of the positive MR and frustrated long range magnetic ordering on the Co-sublattice with spin-glass-like behavior at low temperatures. The $\chi(T)$ data for 340 < T < 380 K also appear to be Curie-Weiss-like and indicate a larger $\mu_{eff}$ compared to the paramagnetic region below $T_{MI}$. However, the small temperature range (~ 40 K) makes it impossible to reasonably accurately determine the effective moment of EuBaCo$_2$O$_{5.5}$ above T$_{MI}$ through a Curie-Weiss fit.

**Conclusions**

Our results suggest an equal number of LS and IS Co$^{3+}$ spins states at least in GdBaCo$_2$O$_{5.5}$ for temperatures below T$_{MI}$. The IS Co sites ferromagnetically order within planes of CoO$_5$ square pyramids, while the LS Co ions correspond to the octahedral sites. The CoO$_6$ participate in the interplane coupling between adjacent CoO$_5$ planes. This coupling is weakly ferromagnetic above T$_N$ and antiferromagnetic below T$_N$. A moderate magnetic field (< 3 T) is sufficient to induce an AFM→FM transition for 150 < T < 260 K. The differences in the magnetic and transport properties of GdBaCo$_2$O$_{5.5}$ and EuBaCo$_2$O$_{5.5}$ are possibly due to a small valence admixture in the nonmagnetic Eu ions, which can play a significant role in the local distribution of the electrons.

**Acknowledgements**

The work in Minsk was supported in part by INTAS grant No. 01-0278 and by SNSF in the frame of the SCOPES program, grant No. 7BYPJ65732. The work at the National High Magnetic Field Laboratory has been supported by the National Science Foundation through Cooperative Agreement No. DMR-0084173 and the State of Florida. PS acknowledges the support by NSF and DOE under grants No. DMR01-05431 and DE-FG02-98ER45797. Both the Minsk and Tallahassee teams gratefully acknowledge the support by the NATO-linkage program through grant PST CLG 979369.

**Figure captions**

FIG. 1: (a) Field cooled M(T) curves of GdBaCo$_2$O$_{5.5}$ at 1000 G for H ⊥ and ∥ c. (b) Field cooled and zero field cooled M(T) curves of EuBaCo$_2$O$_{5.5}$ at 1000 G for H ⊥ and ∥ c.

FIG.2: Isothermal magnetization curves with hysteresis loops for $GdBaCo_2O_{5.5}$ measured at 5 K for $H \perp c$ and $H \parallel c$

FIG. 3: (a) Isothermal magnetization curves of $GdBaCo_2O_{5.5}$ measured at 170 K for $H \perp c$ and $H \parallel c$. Inset: $M'(H) = M(H) - m H$, where the Gd contribution, m, was obtained from the slope of a linear fit to the M(H) data for H > 3 T. (b) Isothermal magnetization curves of $GdBaCo_2O_{5.5}$ measured at 78 K for $H \perp c$, with the Gd contribution subtracted (using the Brillouin function). Inset: Linear plot of the saturation moment vs. $T^3$ for 78 < T <260 K. (c) Isothermal magnetization curves of $EuBaCo_2O_{5.5}$ measured at 5 K for $H \perp c$ and $H \parallel c$.

FIG.4: (a) Resistivity vs. T for $GdBaCo_2O_{5.5}$ measured in the a-b plane and along the c axis. (b) $\ln(\rho)$ vs. 1/T for $GdBaCo_2O_{5.5}$ with current flow in the a-b plane and long the c-axis; simple activation law fit of the a-b plane resistivity data for $GdBaCo_2O_{5.5}$. (c) Low temperature resistivity data for $GdBaCo_2O_{5.5}$ fit to the variable range hopping model.

FIG.5: Resistivity vs. T for $EuBaCo_2O_{5.5}$ measured in the ab plane and along the c axis. Inset: Low temperature resistivity data ($I \parallel c$) for $EuBaCo_2O_{5.5}$ fit to the variable range hopping model for 15 K < T< 120 K.

FIG.6: (a) MR(%) vs. H for $GdBaCo_2O_{5.5}$ at 40 K for $H \perp c$, $H \parallel c$, and (b) MR(%) vs. H for $EuBaCo_2O_{5.5}$ at 40, 78 and 120 K for $H \perp c$ (the arrows indicate the direction of field sweep). (c) MR(%) vs. H at 160 K for $EuBaCo_2O_{5.5}$ for various directions of I and H with respect to the c axis.

\* Present address: Lawrence Livermore National Laboratory, Livermore, CA 94550

\*\* Department of Physics, Florida State University, Tallahassee, FL 32306

Fig 1.

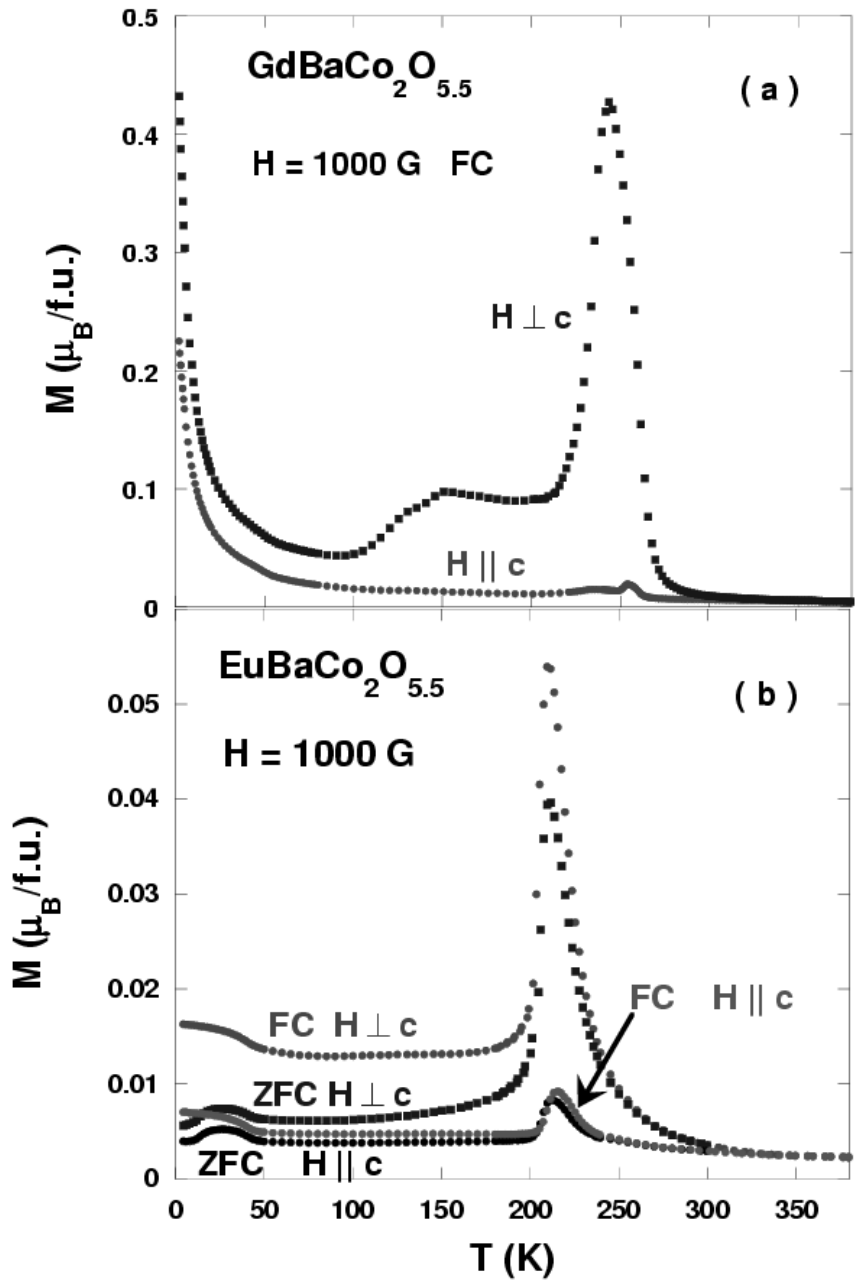

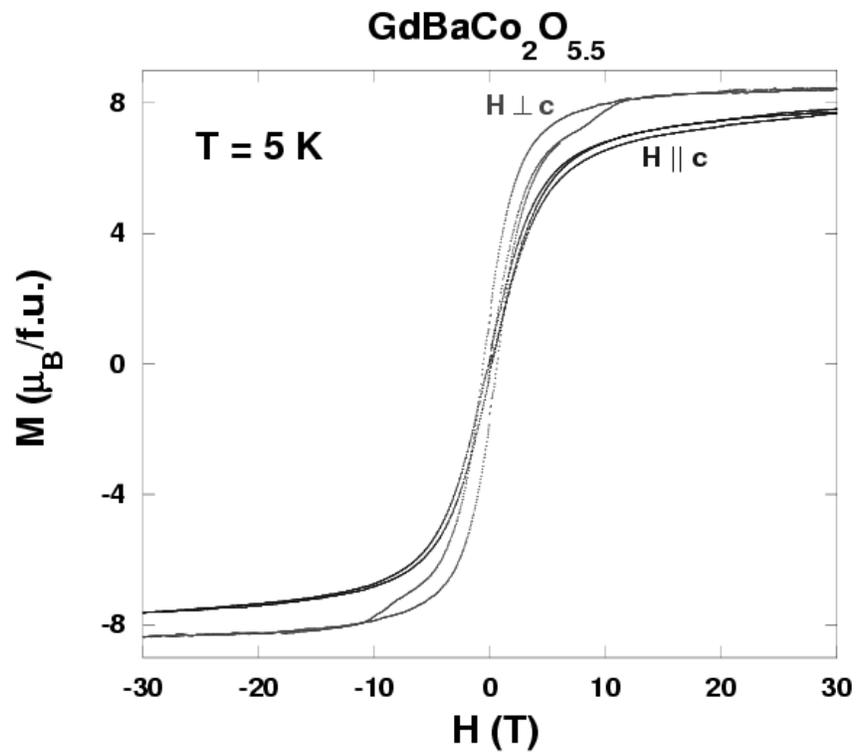

Fig3 a

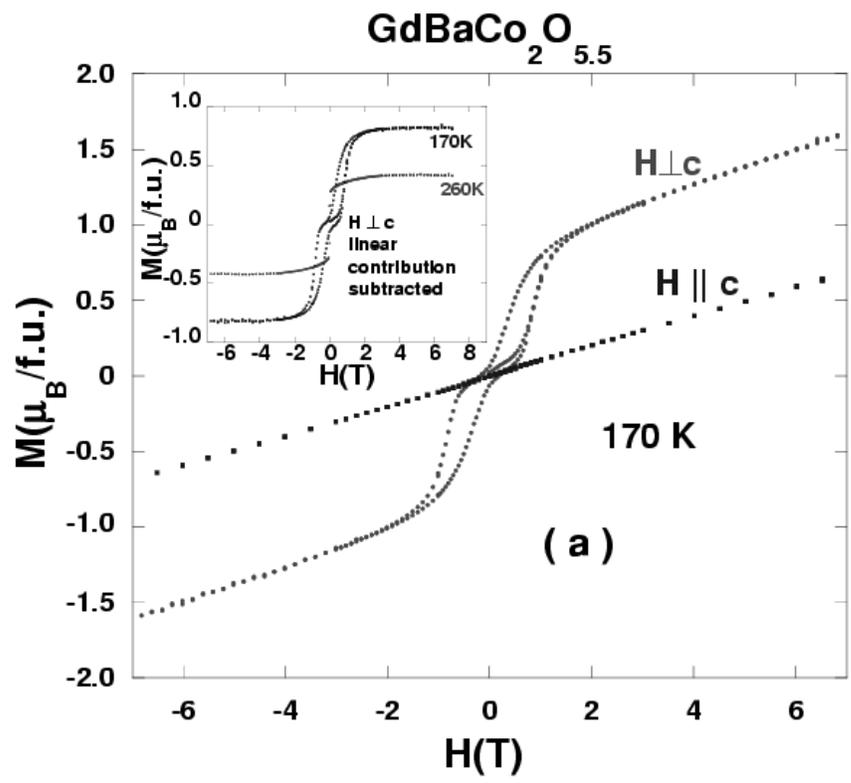

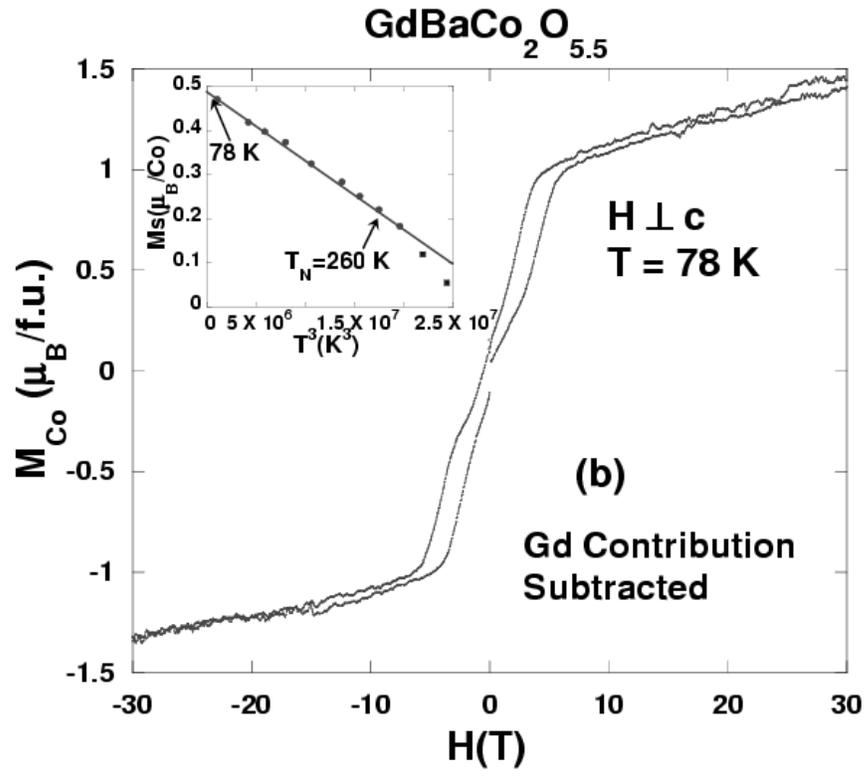

Fig 3. c

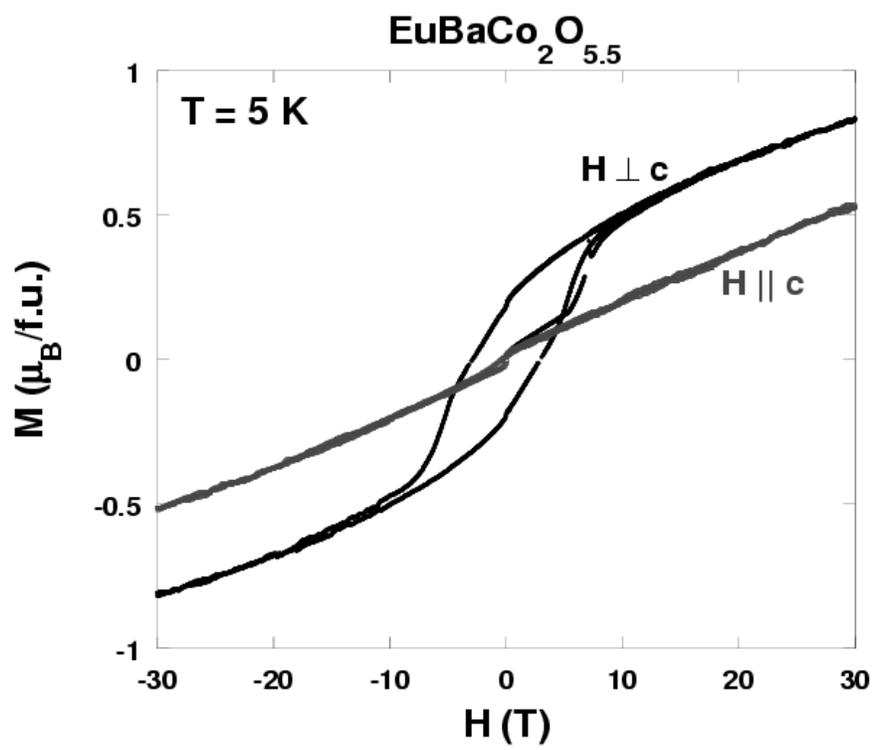

Fig. 4.a

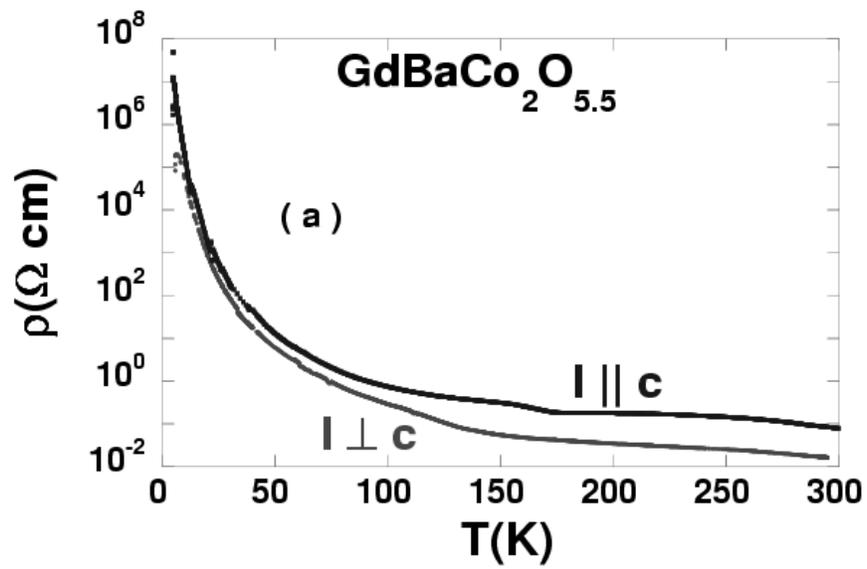

Fig. 4.b

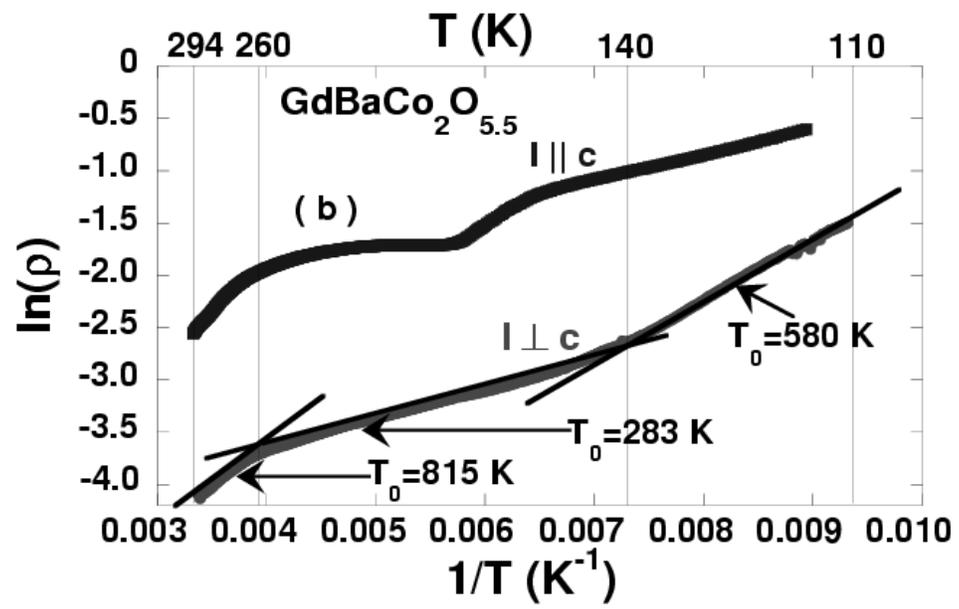

Fig. 4.c

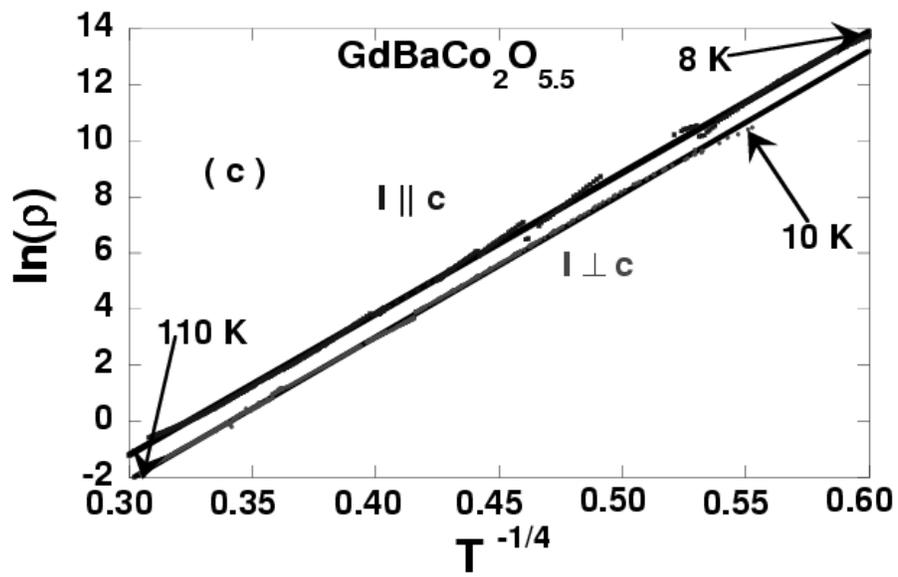

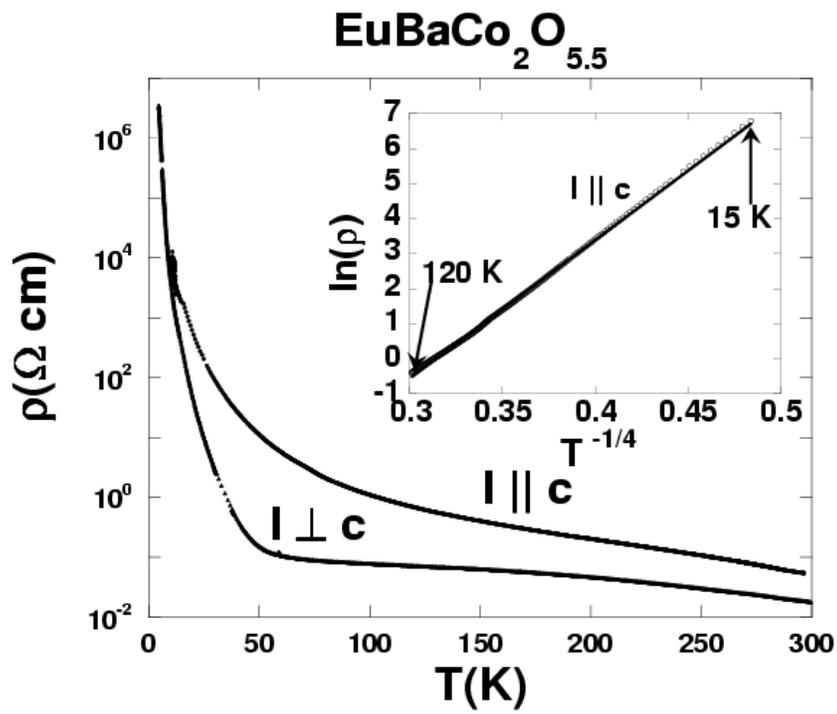

Fig. 5

Fig 6.a

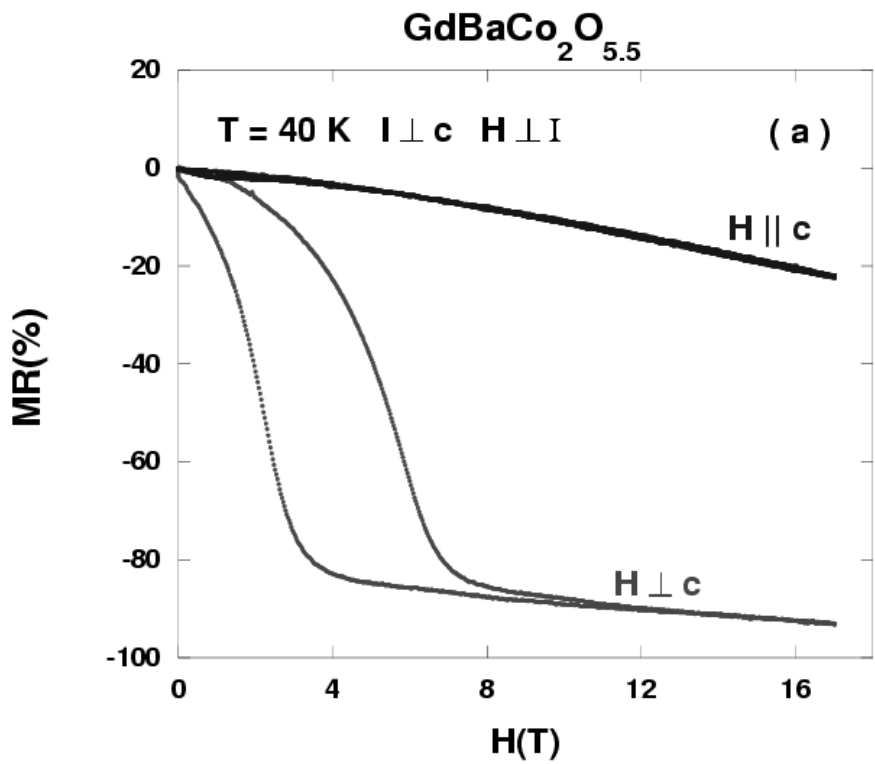

Fig 6.b

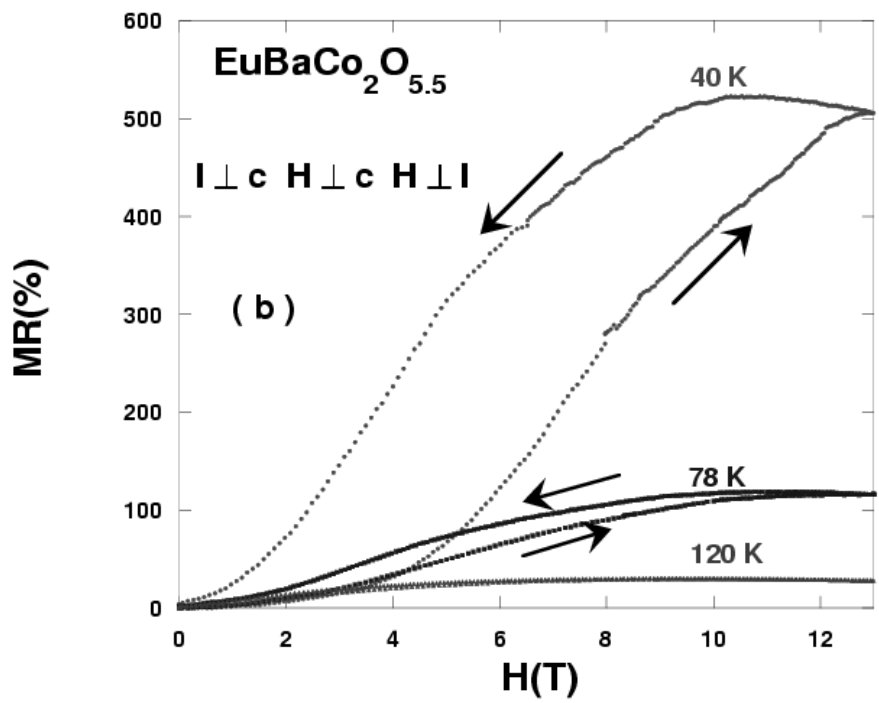

Fig. 6 c

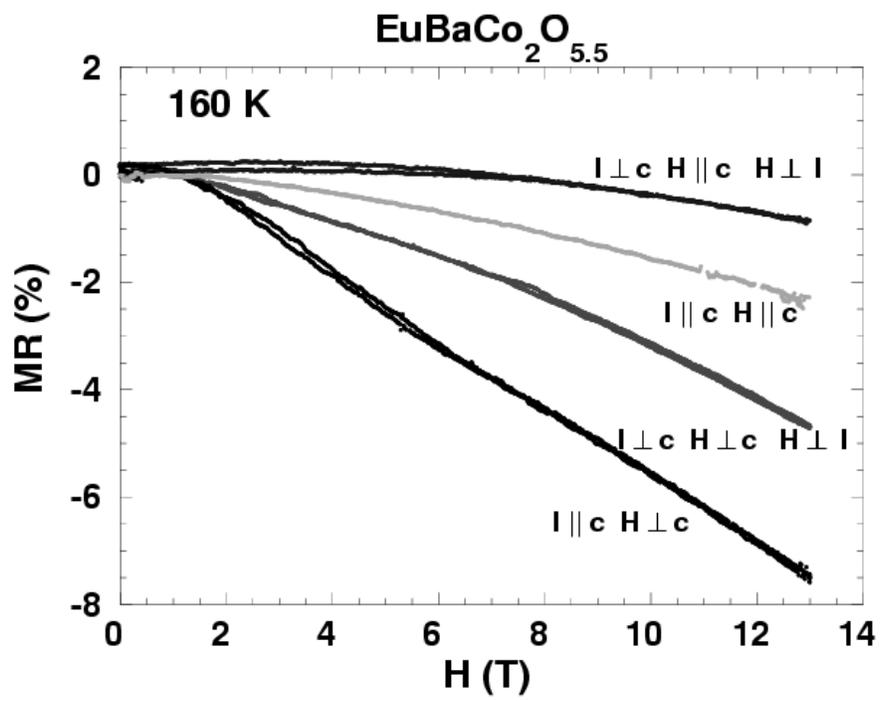